\documentstyle[aps,prl,preprint,psfig,amsmath]{revtex}
\begin{document}

\input{epsf}

\title{Bose-Einstein Condensation at a Helium Surface}
\author{E. W. Draeger$^\dag$ and  D. M. Ceperley}
\address{Department of Physics and National Center for Supercomputing
Applications, University of Illinois Urbana-Champaign, 61801}
\maketitle

\begin{abstract}
Path Integral Monte Carlo was used to calculate the Bose-Einstein
condensate fraction at the surface of a helium film at $T=0.77~K$, as
a function of density.  Moving from the center of the slab to the
surface, the condensate fraction was found to initially increase with
decreasing density to a maximum value of 0.9 before decreasing.  Long
wavelength density correlations were observed in the static structure
factor at the surface of the slab.  Finally, a surface dispersion
relation was calculated from imaginary-time density-density
correlations.
\end{abstract}

\section{Introduction}

It has been suggested\cite{griffin96} that the condensate fraction
in the low density region near a $^4$He surface is significantly
larger than the value in bulk helium of 0.1\cite{penrose56}.
Variational Monte Carlo (VMC) simulations by Lewart and
Pandharipande\cite{lewart88} of small ($N=70$) $^4$He droplets
using a Jastrow one-body (JOB) trial wave function give evidence
for a condensate fraction which approaches unity as the density
goes to zero in the helium surface. However, subsequent
calculations by Galli and Reatto\cite{galli98} have shown that the
condensate fraction throughout a helium surface computed using VMC
is highly sensitive to the choice of trial wave function. They
found that calculations performed using a shadow wave function
with a glue term (G-SWF) have enhanced density-density
correlations at long wavelengths\cite{galli00}, and a maximum
condensate fraction of only 0.5.  Significant ripplon excitations
are expected to reduce the condensate fraction at the surface, as
fewer atoms are able to occupy the zero-momentum state. Quantum
evaporation experiments\cite{wyatt98} can be interpreted as
providing evidence of an enhanced condensate fraction.

To avoid the problem of trial function bias and to include
finite-temperature effects, we have used Path Integral Monte Carlo
(PIMC) to calculate the density-density correlation functions and
condensate fraction at the surface of liquid $^4$He.  We have also
used imaginary-time correlation functions to calculate the
dispersion relation of surface excitations in a free helium
surface, and find good agreement with experimental thin film
measurements.

\section{Simulation Details}

Our simulation system consisted of $^4$He atoms, interacting
pair-wise with a very accurate potential\cite{aziz95}. Periodic
boundary conditions were used with a box size and initial
conditions chosen to favor a double-sided film oriented
perpendicular to the z-axis. To maintain a stable film and
minimize finite size effects, we added an external potential
determined from the long-ranged part of the interaction potential
and the missing atoms from the other side of the slab, so that
atoms on each of the two surfaces saw a potential as if they were
at the surface of a semi-infinite slab\cite{wagner94}.

Most of the calculations were performed with $T=0.77$~K, with an
imaginary time step of $\tau=1/20$~K$^{-1}$.  We performed
simulations of helium slabs containing $N=216$ and $N=432$ atoms,
with dimensions $24\times24\times17$~\AA\ and
$34\times34\times17$~\AA, respectively.  Lower temperature
calculations were also done to determine the temperature
dependence.

\section{The Static Structure Factor}
\label{section:slabskz}

In order to determine the extent to which ripplons are present in
a free helium surface, we estimated density-density correlations
at the surface\cite{galli00} with the static structure factor
defined as:
\begin{equation}
S(k_{||};z,z') = \langle \rho_{k_{||}}(z) \rho_{k_{||}}(z')\rangle
\end{equation}
\noindent where $\rho_k(z)\equiv\frac{1}{\sqrt{N(z)}}\sum\limits_i
e^{i{\bf k}\cdot{\bf r}_i}\delta(z_i-z)$, $N(z)$ is the number of
particles in the bin at position $z$ and $k_{||}$ is the wave
vector parallel to the surface.  This measures the correlation
between density fluctuations at vertical positions $z$ and $z'$.

Fig.~\ref{slabskz} shows $S(k_{||},z,z)$ curves as a function of
density $\rho(z)$.  Each curve is the average of both sides of the
slab of eight identical simulations (a total of $8\times 10^5$ Monte
Carlo passes).  At $T=0.77$~K, for densities of
$\rho(z)=0.015$~\AA$^{-3}$\ and below there is a small enhancement of
long-wavelength density-density correlations at $k_{||} = 2\pi/L =
0.18$~\AA$^{-1}$, which is evidence for ripplons.  However, the curves
are closer to the VMC calculations of Galli and Reatto which used a
JOB trial wave function than those which used the G-SWF form.  This
does not imply that the JOB trial wave function is well-suited to
representing an inhomogeneous helium system such as a helium slab, but
rather the degree to which the G-SWF significantly overestimates the
effect of ripplons in a free helium surface.  Calculations at $N=216$
and $T=0.77$~K agree with the $N=432$, $T=0.77$~K results of
Fig.~\ref{slabskz} within statistical error, indicating that
finite-size effects are negligible.  At $T=0.38$~K and $N=216$, we
find a measurable decrease in the long-wavelength correlations at
$k_{||}=2\pi/L = 0.26$~\AA$^{-1}$.  Further studies are needed to establish the
temperature dependence.

\section{The Excitation Spectrum}

The excitation spectrum can be estimated with path integrals using
imaginary-time correlation functions\cite{ceperley95}.  The
dynamic structure factor is related to the imaginary-time
density-density correlation function by:
\begin{align}
F({\bf k},t) & = \int\limits_{-\infty}^{\infty} d\omega \; e^{-t
\omega} \; S({\bf k},\omega) \label{fkt1}
 \\
& = \frac{1}{N}\left\langle
\rho_{\bf k}(t) \rho_{\bf k}(0) \right\rangle.
\label{fkt2}
\end{align}
To select out the excitations at the free surface, we want to
calculate the imaginary-time correlation function of propagating
surface modes.  W. F. Saam\cite{saam75} proposed that the lowest quantized
hydrodynamic mode (capillary wave) at a free helium surface will have
the form
\begin{equation}
\phi_{k0}({\bf r},t) = \phi_{k0}(z)\, e^{i {\bf k}\cdot{\bf r}_{||}}
\, e^{-i \omega_{k0} t}
\end{equation}
where
\begin{equation}
\phi_{k0}(z) \propto e^{-\kappa(k) z}
\end{equation}
and the decay constant $\kappa(k)$ is defined as
\begin{align}
\label{fkt3}
& \kappa(k) = -b_k + (k^2+b_k^2)^{1/2} \\
& b_k \equiv \frac{\sigma_0 k^2}{2\rho_0s^2}
\end{align}
where $\sigma_0$ is the zero-temperature surface tension, $\rho_0$ is
the bulk density, and $s$ is the zero-temperature sound velocity.  To
calculate the dispersion relations for excitations of this form, we
use $\tilde{\rho}_{\bf k}$ in Eq.~(\ref{fkt2}), defined as
\begin{equation}
\tilde{\rho}_{\bf k} = \sum\limits_i \; e^{i {\bf k}_{||}\cdot{\bf r}_{i||}} \; \phi_k(z_i)
\end{equation}
where
\begin{equation}
\phi_k(z) =
\begin{cases}
e^{-\kappa(k)(z-z_c)} & \text{if $z\leq z_c$,} \\
1 & \text{if $z > z_c$} \\
\end{cases}
\end{equation}
and $\kappa(k)$ is defined by Eq.~(\ref{fkt3}).  We defined $z_c$ as
the point in the surface at which the average density was equal to
10\% of the bulk density.  Tests have shown that the results are not
sensitive to the value of $z_c$.

Extracting the dynamic structure factor by inverting
Eq.~(\ref{fkt1}) is ill-conditioned in the presence of statistical
noise.  It has been shown\cite{gubernatis91} that a maximum
entropy method greatly increases the numerical stability.  In the
maximum entropy method, $S(k,\omega)$ is calculated by minimizing
the function
\begin{equation}
{\cal F}(S,\alpha) = \frac{e^{-(1/2)Q(S)}}{Z_Q} \times \frac{e^{\alpha\,
\zeta(S)}}{Z_{{\cal \zeta}}}
\end{equation}
where $Q(S)$ is the ''likelihood'' of the PIMC data given an $S$
and $\zeta(S)$ is the entropy of a given $S(k,\omega)$ defined
with respect to some default model, with $\alpha$ an adjustable
parameter (also optimized).

The dispersion energy for a given value of $k$ can be determined from
the position of the maximum value of $S(k,\omega)$.  Boninsegni and
Ceperley\cite{boninsegni96} found that the position of the main
$S(k,\omega)$ peaks for liquid helium agree quite well with
experiment, despite significant broadening of the excitation spectrum
caused by the maximum entropy procedure.  The dispersion energy of the
surface excitations as estimated using this procedure are shown in
Fig.~\ref{ripplon_disp}.  The two lowest data points, at $k <
0.5$~\AA, had significant fitting error and are only qualitatively
reliable.  Otherwise, we see excellent agreement with the experimental
thin film data of Lauter \emph{et al.}\cite{lauter92}, including the
curvature of the ripplon branch toward the roton minimum, proposed as
evidence for roton-ripplon
hybridization\cite{krotscheck87,pitaevskii92}.

\section{The Bose-Einstein Condensate Fraction}

We define the condensate fraction in the slab geometry by the
fraction of atoms at a given value of $z$ having precisely
$k_{||}=0$. (Because $[k_{||},z]=0$, we can measure the momentum
parallel to the surface simultaneously with the z-position.)  The
momentum distribution at a distance $z_0$ from the center of the
slab is given by
\begin{equation}
n_{\bf k_{||}}(z_0) = \frac{1}{(2\pi)^2}\int d{\bf r_{||}} \;
e^{-i {\bf k_{||}}\cdot{\bf r_{||}}} \; n({\bf r_{||}};z_0)
\end{equation}
where the off-diagonal single particle density matrix is:
\begin{align}
n
({\bf r};z) = & \frac{1}{\rho(z)Z}\int d{\bf r}_1 \cdots d{\bf r}_N \nonumber \\
& \times \rho({\bf r}_1,{\bf r}_2,\cdots {\bf r}_N,{\bf r}_1+{\bf
r},{\bf r}_2,\cdots {\bf r}_N;\beta),
\label{nofr1}
\end{align}
\noindent
where $\rho$ is the many-body density matrix and $Z={\rm Tr}(\rho)$ is
the partition function.  This function can be calculated from
PIMC\cite{ceperley87} by performing simulations with a single open
path. We fix the endpoints of the open path at $z=z_0$, and
calculate the distribution of end-to-end distance $n(r_{||};z_0)$.
The condensate fraction at a given point in the surface is:
\begin{equation}
n_0(z_0) =  \frac{n(r_{||}\rightarrow\infty;z_0)}{n(r_{||}\rightarrow 0;z_0)}.
\label{slabcondfract}
\end{equation}

Using PIMC, we calculated $n(r_{||},z)$ throughout the slab.
Nonlinear least-squares fitting was used in the region of $r_{||} <
1.5~\AA$ to get an estimate of $n(r_{||}=0,z)$, and
$n(r_{||}\rightarrow\infty,z)$ was calculated by averaging over the
region at large $r_{||}$ where $n(r_{||},z)$ is flat. At the lowest
densities, it is not clear whether $n(r)$ has reached its asymptotic
limit within the finite simulation box.  Thus, for $N=216$ at
densities below $\rho(z)=0.001$~\AA$^{-3}$ our results are upper
bounds to the condensate fraction.

The condensate fraction $n_0(z)$ is plotted as a function of average
density $\rho(z)$ in Fig.~\ref{slabnofr}, for both $N=216$ and $N=432$
helium slabs.  As one moves through the surface, the condensate
fraction initially increases with decreasing density, due to the
decreased zero-point motion from helium-helium interactions, reaching
a maximum value of 0.93(3) at $\rho(z) = 0.002$~\AA$^{-3}$.  As the
average density decreases below this point, the condensate fraction
begins to decrease, further evidence for correlated density
fluctuations due to ripplons at the surface.  This is in qualitative
agreement with the G-SWF VMC calculations of Galli and Reatto.
However, the G-SWF trial wave function significantly overestimates the
degree to which ripplons are present in the surface.

We find that the probability of a given atom belonging to a long
exchange cycle does not change if that atom is located above the
surface.  Hence the density fluctuations pulling atoms above the
surface are accompanied by other (exchanging) atoms, in qualitative
agreement with the ripplon model.  But there is also present an
appreciable fraction ($\approx30\%$) of non-exchanging atoms, possibly
a result of finite temperature excitations.

We have presented PIMC calculations of the density-density correlation
functions and condensate fraction at a free helium surface.  These
results support the model of a free helium surface with ripplons, in
which the condensate fraction reaches a maximum at an intermediate
density in the liquid-vacuum interface, before decreasing at lower
densities. Experimental probes of the surface will indeed see an
enhanced condensate fraction as proposed by Griffin and
Stringari\cite{griffin96}.

\begin{acknowledgements}
This research was carried out on the Origin 2000 at the National
Center for Supercomputing Applications and the IBM cluster at the
Materials Computation Center, and was supported by the NASA
Microgravity Research Division, Fundamental Physics Program.  This
work was also performed under the auspices of the U.S. Department of
Energy by University of California Lawrence Livermore National
Laboratory under contract No. W-7405-Eng-48.

$\dag$  Present Address:  Lawrence Livermore National Laboratory, 7000 East Avenue, L-415, Livermore, CA  94550.

\end{acknowledgements}

\begin{figure}[hbt]
\caption{PIMC density distribution, for both $N=216$ and $N=432$,
$T=0.77$~K semi-infinite $^4$He slabs (open and solid circles).  The
two curves are indistinguishable.  The dashed line shows the effective
helium density felt by atoms at $z>0$.}
\label{slabden}
\end{figure}

\begin{figure}[hbt]
\caption{$S(k_{||},z,z)$ vs. $k_{||}$ throughout the surface region in
an $N=432$, $T=0.77$~K slab, calculated with PIMC.}
\label{slabskz}
\end{figure}

\begin{figure}[hbt]
\caption{Dispersion relation of free surface excitations.  The results
calculated from PIMC and maximum-entropy inversion (filled diamonds)
for an $N=216$, $T=0.77$~K $^4$He slab are compared with the
experimental thin film data (open circles) of Lauter \emph{et al.} and the
DFT results (solid line) of Lastri \emph{et al.}  Also shown is the
dispersion relation of bulk $^4$He at SVP (dashed line).}
\label{ripplon_disp}
\end{figure}

\begin{figure}[hbt]
\caption{$n(r_{||},z)$ vs $r_{||}$, throughout the surface region of
$N=216$ (top) and $N=432$ (bottom) $^4$He slabs at $T=0.77$~K,
calculated with PIMC.  Density labels correspond to
$z=0,8,9,10,10.5,11,12,13$~\AA\ for $N=216$, and $z=0,9,11,12,13$~\AA\
for $N=432$.}
\label{nofrcurves}
\end{figure}

\begin{figure}[hbt]
\caption{Condensate fraction vs. density throughout the surface region
of a $T=0.77$~K $^4$He slab, calculated from the $n(r_{||},z)$ curves
in Fig.~\ref{nofrcurves}.  Also shown are the VMC calculations of
Galli and Reatto.}
\label{slabnofr}
\end{figure}

\pagebreak

\begin{center}
\epsfxsize=\columnwidth
\centerline{\epsfbox{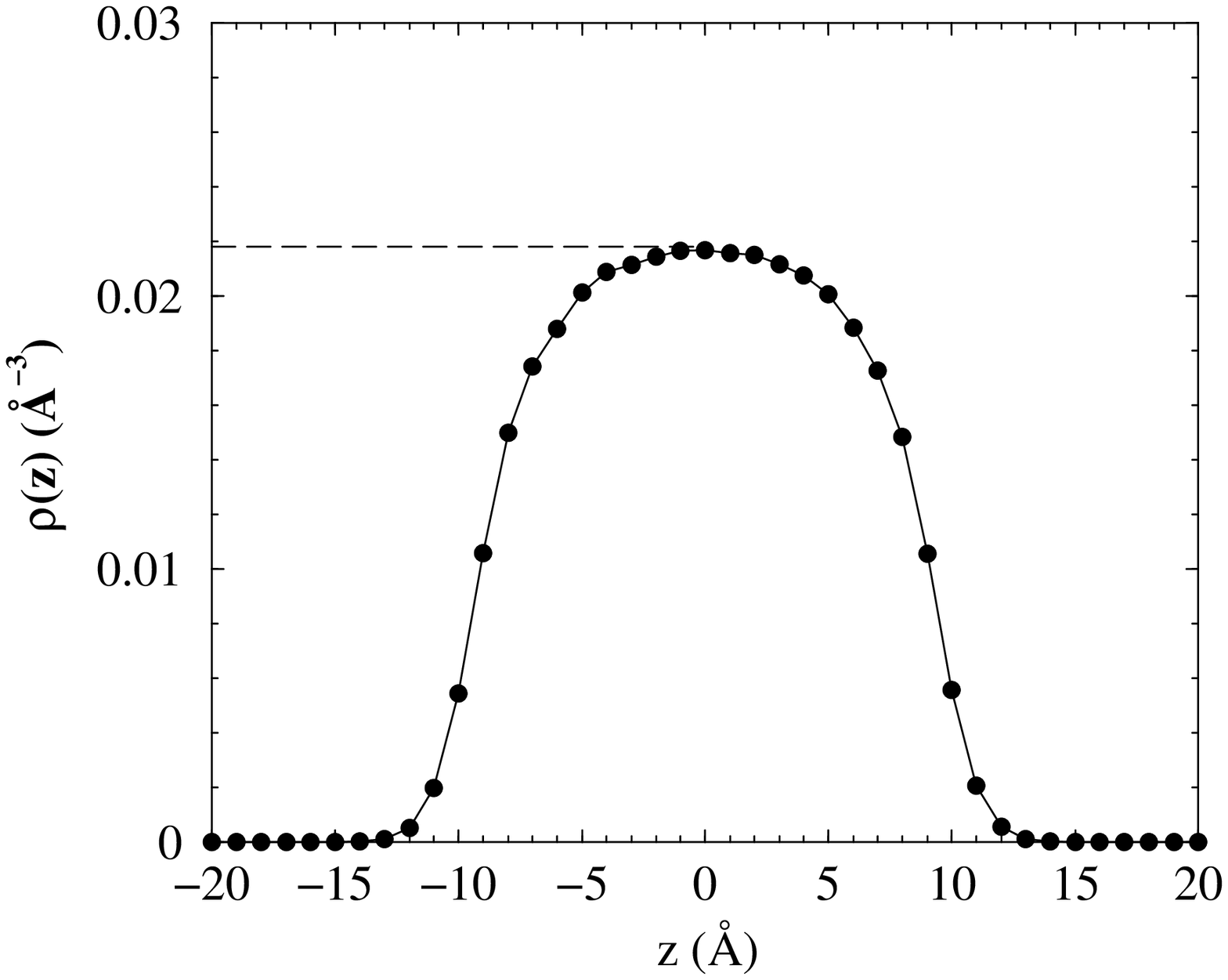}}
\end{center}
\vfill
Draeger Figure 1

\begin{center}
\epsfxsize=\columnwidth
\centerline{\epsfbox{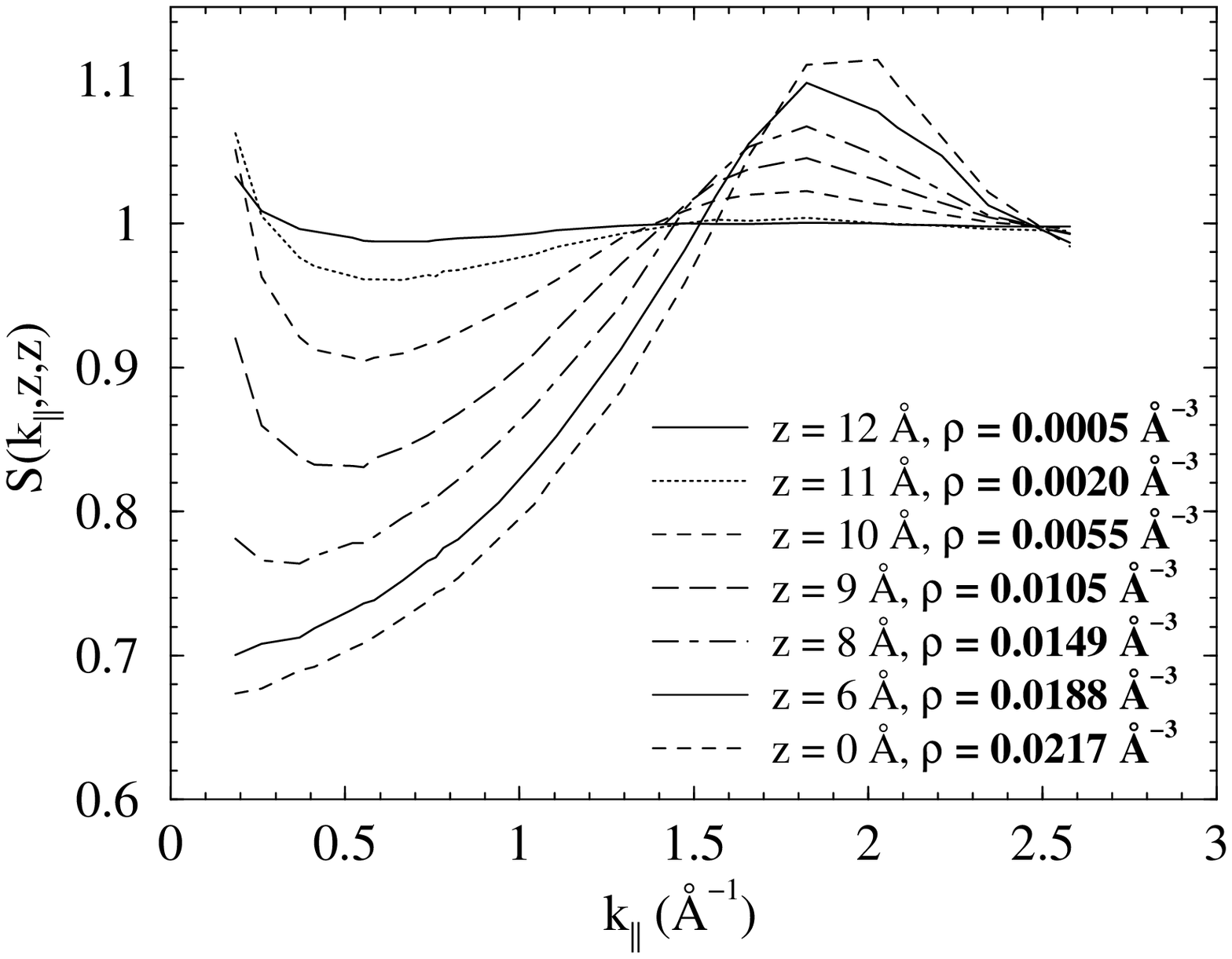}}
\end{center}
\vfill
Draeger Figure 2

\begin{center}
\epsfxsize=\columnwidth
\centerline{\epsfbox{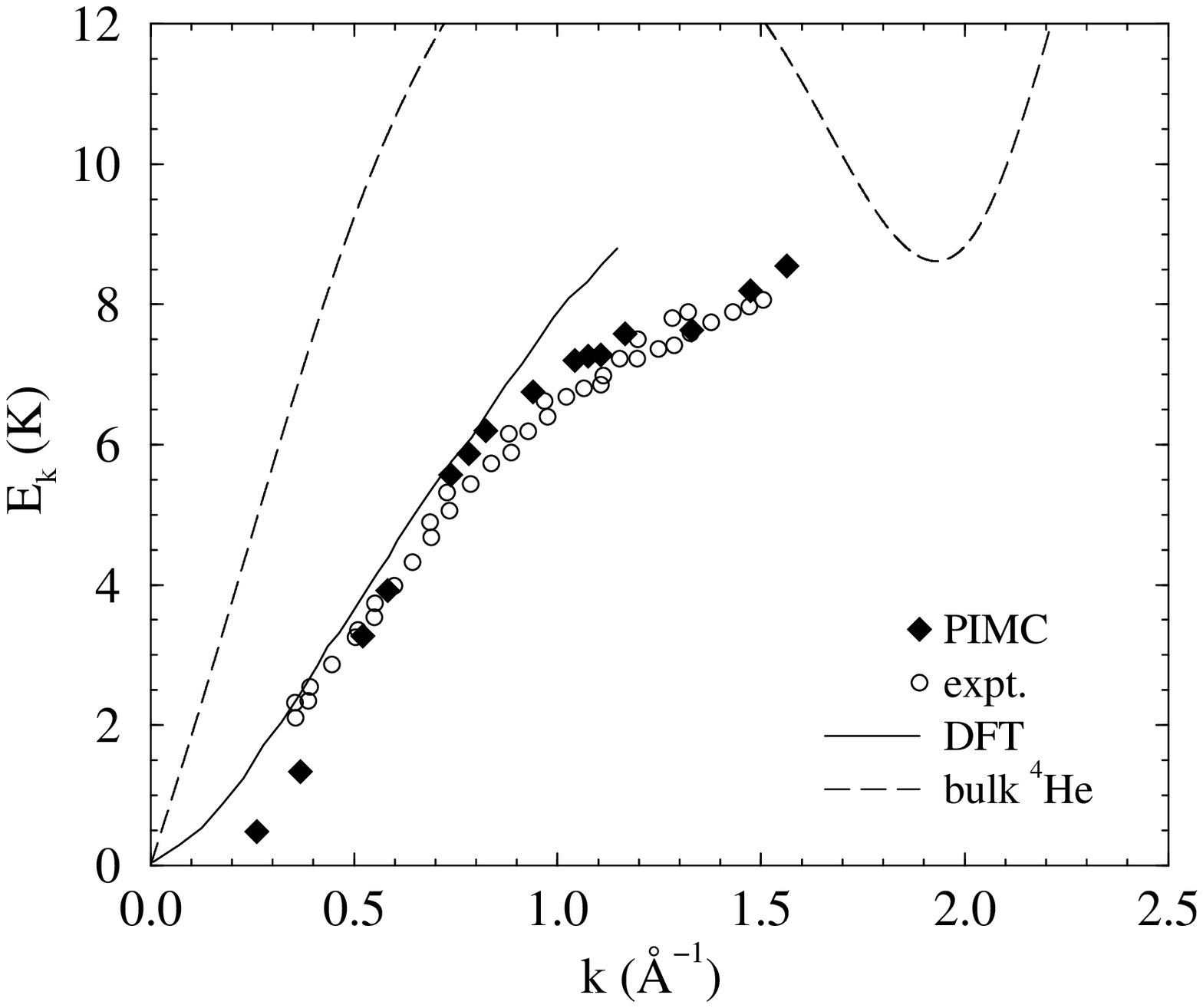}}
\end{center}
\vfill
Draeger Figure 3

\begin{center}
\epsfxsize=4.8in
\centerline{\epsfbox{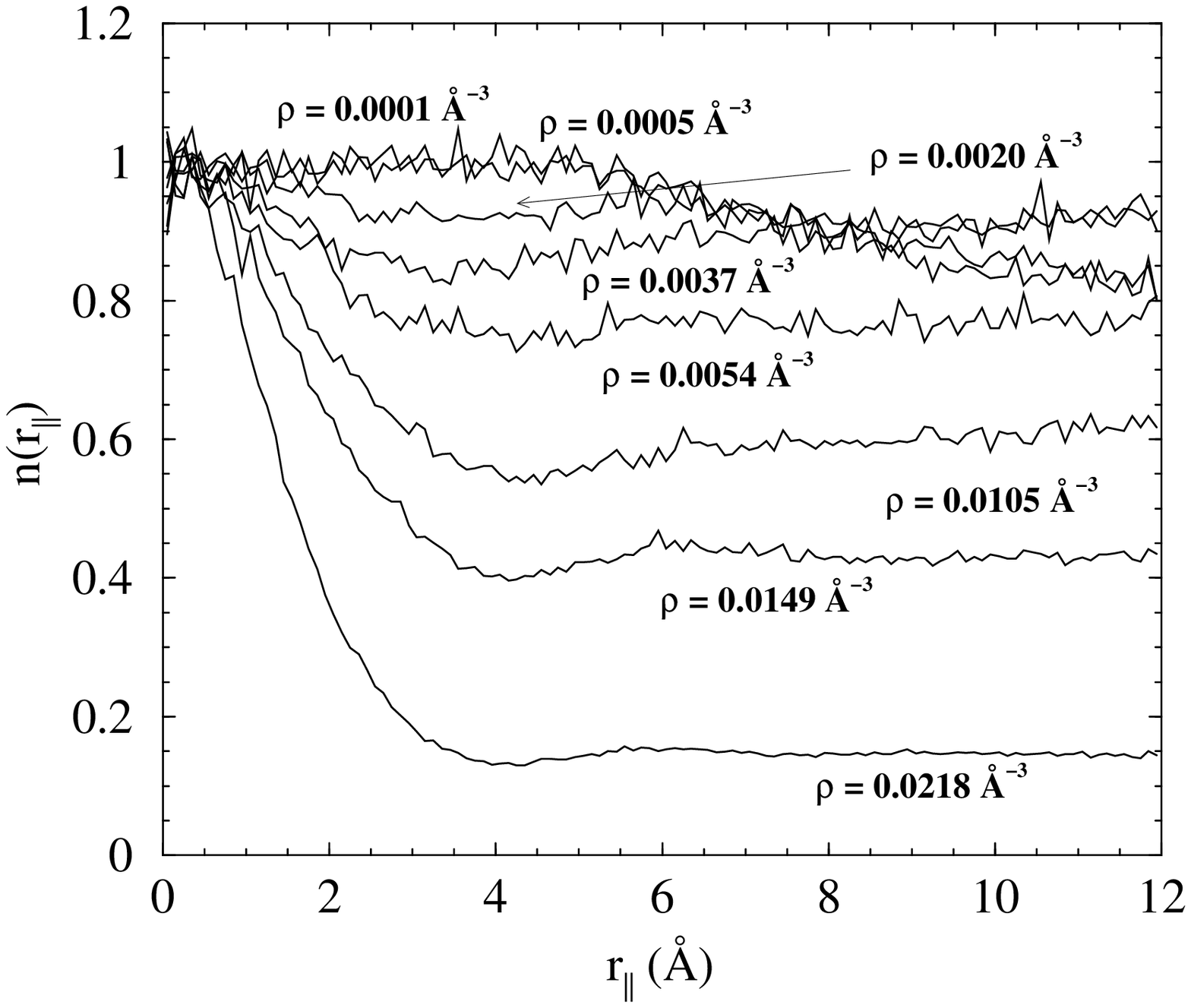}}
\epsfxsize=4.8in
\centerline{\epsfbox{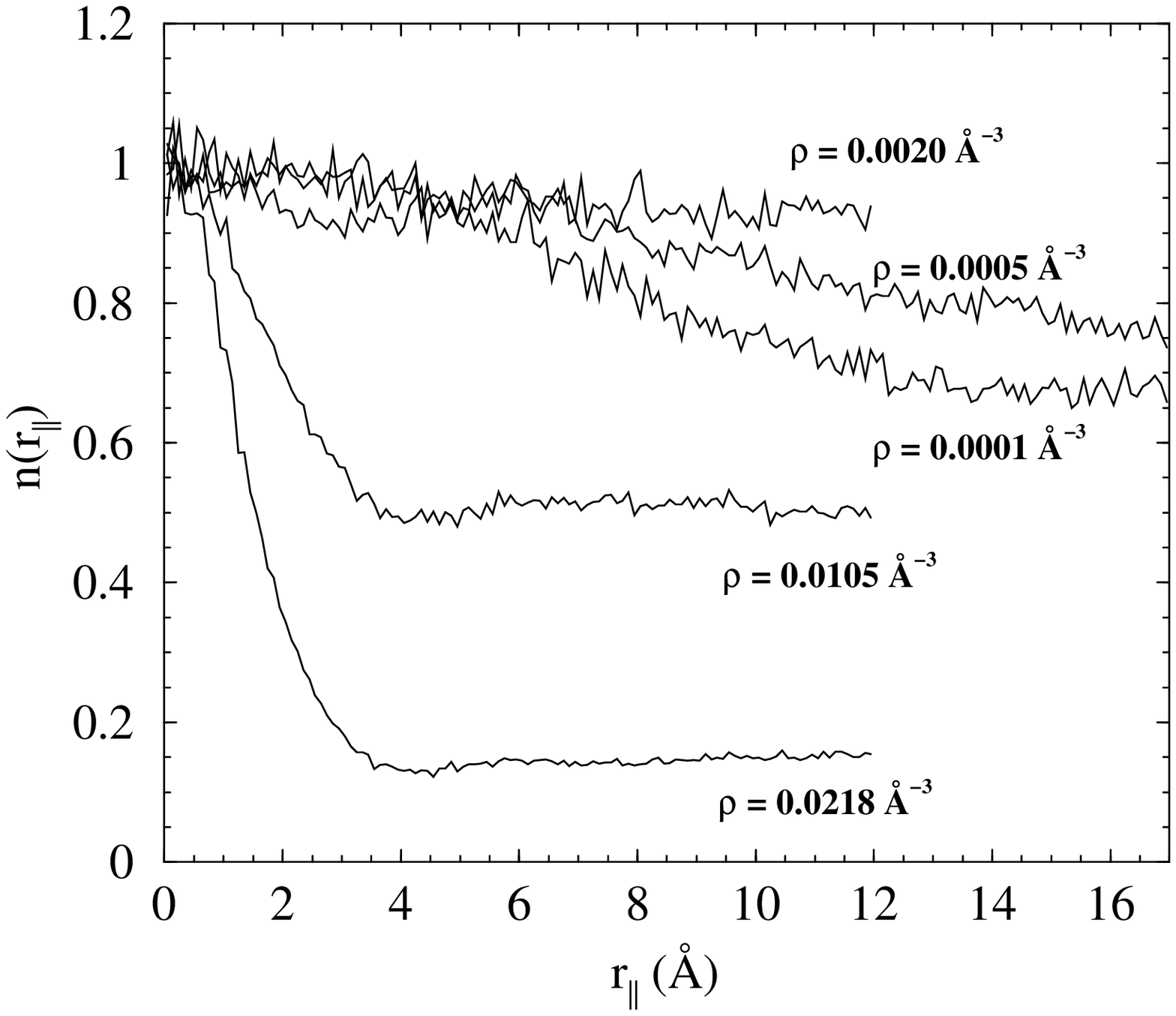}}
\end{center}
\vfill
Draeger Figure 4

\begin{center}
\centerline{\epsfbox{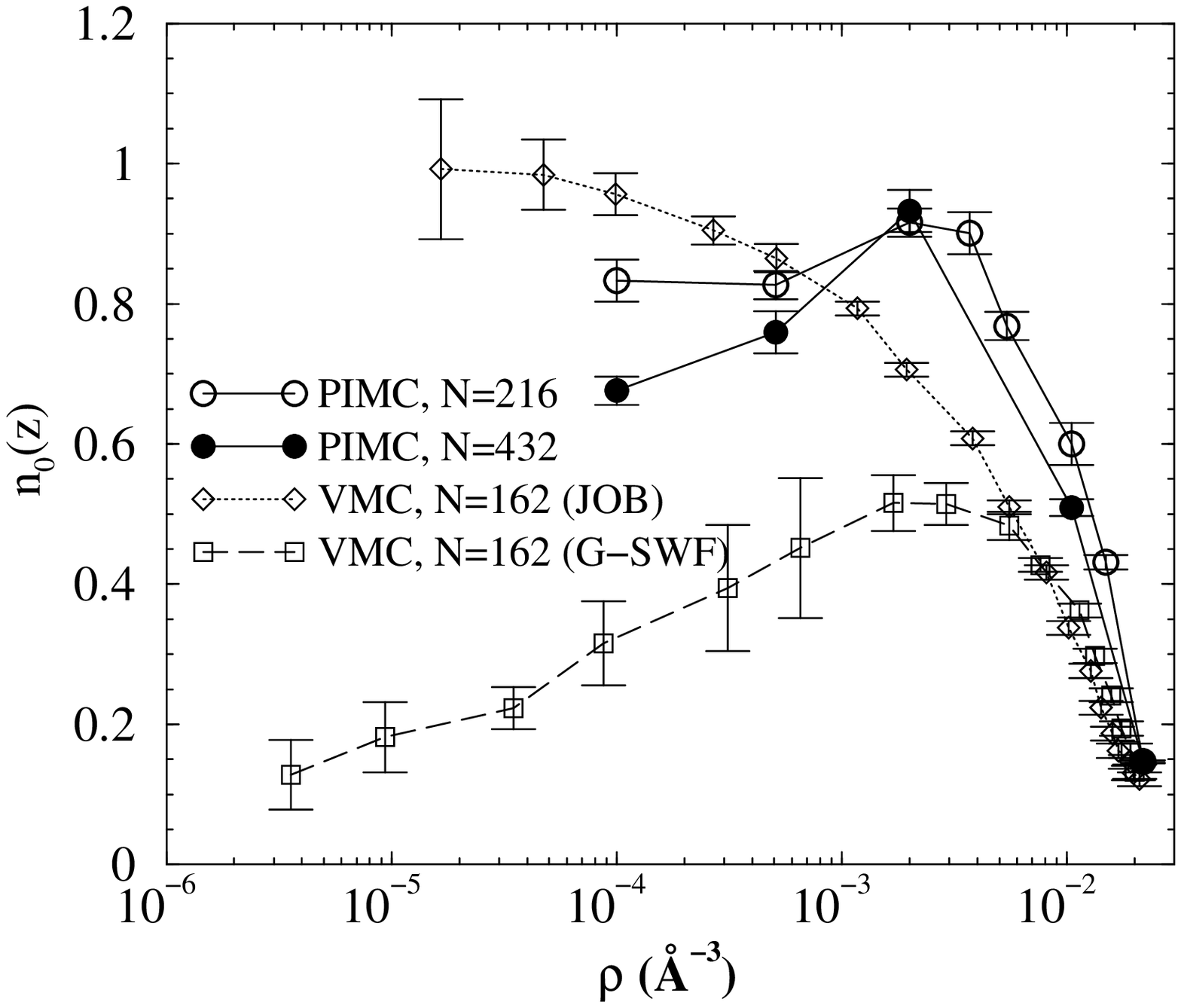}}
\end{center}
\vfill
Draeger Figure 5

\end{document}